\documentclass[acmsmall, nonacm]{acmart}
\setcopyright{none}

\AtBeginDocument{%
  }

\usepackage{soul} % for the command \hl
\usepackage{float}

\usepackage{algorithm}
\usepackage{algpseudocode} % or algorithmic

\usepackage{tabularx}
\usepackage{natbib}
\usepackage{subcaption}

\begin{document}

%% Note: You usually need to keep \maketitle here
\title{Lattice Surgery Aware Resource Analysis for the Mapping and Scheduling of Quantum Circuits for Scalable Modular Architectures}

\author{Batuhan Keskin}
\authornote{Both authors contributed equally to this research.}
\affiliation{%
 \institution{EPFL}
 \city{Lausanne}
 \country{Switzerland}}
\email{batuhan.keskin@epfl.ch}
\orcid{0009-0002-6464-5476}

\author{Cameron Afradi}
\authornotemark[1]
\affiliation{%
 \institution{University of California, Berkeley}
 \city{California}
 \country{USA}}
\email{cameronafradi@berkeley.edu}
\orcid{0009-0003-1262-0127}

\author{Sylvain Lovis}
\affiliation{%
 \institution{EPFL}
 \city{Lausanne}
 \country{Switzerland}}
\email{sylvain.lovis@epfl.ch}
\orcid{0009-0005-8182-6617}

\author{Maurizio Palesi}
\affiliation{%
 \institution{University of Catania}
 \city{Catania}
 \country{Italy}}
\email{maurizio.palesi@unict.it}
\orcid{0000-0003-3129-0664}

\author{Pau Escofet}
\affiliation{%
 \institution{Universitat Politècnica de Catalunya}
 \city{Barcelona}
 \country{Spain}}
\email{pau.escofet@upc.edu}
\orcid{0000-0003-3372-1931}

\author{Carmen G. Almudever}
\affiliation{%
 \institution{Universitat Politècnica de València}
 \city{València}
 \country{Spain}}
\email{cargara2@upv.edu.es}
\orcid{0000-0002-3800-2357}

\author{Edoardo Charbon}
\affiliation{%
 \institution{EPFL}
 \city{Lausanne}
 \country{Switzerland}}
\email{edoardo.charbon@epfl.ch}
\orcid{0000-0002-0620-3365}

\renewcommand{\shortauthors}{Keskin et al.}

\begin{abstract}
Quantum computing platforms are evolving to a point where placing high numbers of qubits into a single core comes with certain difficulties such as fidelity, crosstalk, and high power consumption of dense classical electronics. Utilizing distributed cores, each hosting logical data qubits and logical ancillas connected via classical and quantum communication channels, offers a promising alternative. However, building such a system for logical qubits requires additional optimizations, such as minimizing the amount of state transfer between cores for inter-core two-qubit gates and optimizing the routing of magic states distilled in a magic state factory. 

In this work, we investigate such a system and its statistics in terms of classical and quantum resources. First, we restrict our quantum gate set to a universal gate set consisting of CNOT, H, T, S, and Pauli gates. We then developed a framework that can take any quantum circuit, transpile it to our gate set using Qiskit, and then partition the qubits using the KaHIP graph partitioner to balanced partitions. Afterwards, we built an algorithm to map these graphs onto the 2D mesh of quantum cores by converting the problem into a Quadratic Assignment Problem with Fixed Assignment (QAPFA) to minimize the routing of leftover two-qubit gates between cores and the total travel of magic states from the magic state factory. Following this stage, the gates are scheduled using an algorithm that takes care of the timing of the gate set. As a final stage, our framework reports detailed statistics such as the number of classical communications, 
the number of EPR pairs and magic states consumed, and timing overheads for pre- and post- processing for inter-core state transfers. These results help to quantify both classical and quantum resources that are used in distributed logical quantum computing architectures.
\end{abstract}


\ccsdesc[500]{Computing methodologies~Quantum computing}
\ccsdesc[300]{Computer systems organization~Dependable and fault-tolerant systems and networks}
\ccsdesc[200]{Software and its engineering~Compilers}

\keywords{lattice surgery; surface code; quantum error correction; resource estimation; quantum compilation; circuit mapping; scheduling; modular quantum architectures; fault-tolerant quantum computing}

\maketitle

\section{Introduction}
\begin{figure}   
    \centering
    \includegraphics[width=0.95\linewidth]{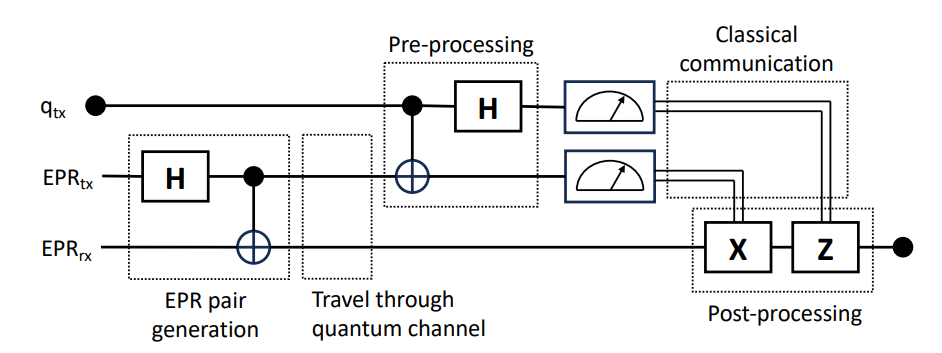}
    \caption{Simple teleportation circuit}
    \label{fig:simpletel}
\end{figure}

Distributed quantum computing (DQC) for quantum information applications is a promising development for overcoming current problems in quantum computers, such as high noise due to large amounts of qubits in close proximity, cross-talk between qubits, and dense control and readout electronics that lay near the qubits \cite{PhysRevApplied.18.044064,9923784,LaRacuente2025modelingshortrange}. Similar to the classical computing domain, future quantum computers will also be integrated with networks, routers, and modular cores \cite{Bravyi_2022}. Such a structure should have the capability to communicate between cores using both classical and quantum channels. Usually, the quantum channel encompasses an EPR pair generator and distributor to implement the circuit in Figure \ref{fig:simpletel}, while the classical channels are used to route the measurement results in the quantum teleportation circuits \cite{GottesmanChuang1999Teleportation}. Significant progress has been made to create these quantum channels using photon-shuttling interconnects, capable of shuttling microwave photons along a superconducting waveguide to facilitate remote entanglement between spatially separated quantum modules \cite{Almanakly_2025} or utilizing low-loss aluminum coaxial interconnects, which have demonstrated inter-module quantum state transfer and Bell state entanglement with fidelities up to 99\% \cite{Niu_2023}. Other methods have proposed using microwave photons for entanglement generation \cite{ang2022architecturesmultinodesuperconductingquantum} for superconducting qubits, entangling gates for neutral atoms using Rydberg-blockade interactions \cite{Evered2023}, and shuttling ions for ion-trapped computers \cite{10.1116/1.5126186}.

For fault-tolerant quantum applications with distributed cores, the structure of these entanglement generation and the layout of the cores must be designed with additional care. Since each core contains logical qubits, the gate operations have to be done without destructing the encoding capability of the qubits both for inter-core and intra-core operations. Considering logical qubits  produced by rotated surface codes, there are two ways of employing logical qubits over a DQC. Logical qubits may be encoded on all/some of the cores by encoding the logical state onto the physical qubits that are in  different cores \cite{sutcliffe2025distributedquantumerrorcorrection}, or alternatively, logical qubits in each of the cores are encoded over some set of physical qubits within the same core. To do the former method, the system has to generate the necessary connection between physical qubits (GHZ state) each time there is a stabilizer check \cite{Ryan-Anderson2024}. In the latter approach, the direction of this work, inter-core operations are only required for logical gates between logical qubits residing on different cores.

Building on this framework, an effective DQC architecture must not only ensure the fault-tolerant execution of logical operations, but also provide scalable mechanisms for routing, scheduling, and supplying essential non-Clifford resources \cite{PhysRevA.71.022316}. In particular, the efficient movement of logical states and the provisioning of magic states across a network of cores is critical for enabling universal computation while keeping both quantum and classical resource overhead manageable. Motivated by these challenges, our work makes the following contributions:

\begin{itemize}
  \item We developed a simulation framework for DQC with logical qubits to analyze classical and quantum resource consumption.
  \item We devised a routing system that uses logical ancillas with logical smooth teleportation gates for state transfer between logical qubits to schedule and route distant CNOTs and magic states.
  \item We introduced a Magic State Factory (MSF) connected to a 2D mesh of quantum cores from one edge to supply magic states for T and S gates, and a new optimization procedure for the qubit-to-core and core-to-mesh placement by taking into account the routing of CNOTs and magic states.
  \item We created a scheduling algorithm that parallelizes and pipelines gates as much as possible according to the availability of logical qubits and timing of logical gates.  
  \item We ran different benchmarks and report the consumed resources for both quantum NoC and classical NoC.
\end{itemize}
\section{Background}
\subsection{Surface Code}
\begin{figure}   
    \centering
    \includegraphics[width=0.25\linewidth]{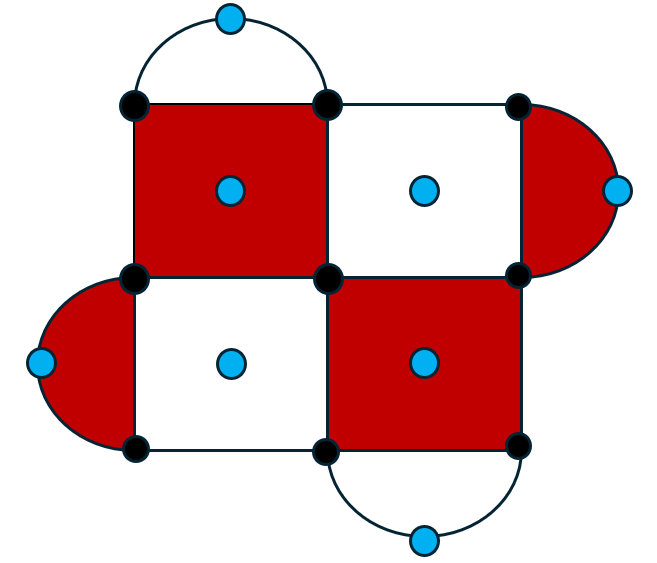}
    \caption{Distance-3 Rotated Surface Code. Black circles are data qubits, blue circles are ancilla qubits. Red and white represent different stabilizer checks such as X and Z.}
    \label{fig:surfacecode}
\end{figure}

It is necessary to first introduce the concepts of surface code and lattice surgery (LS), as they appear frequently in this work. 
The surface code is a topological quantum error correction code \cite{Kitaev_2003}. It is one of the most well-known and studied quantum error correction codes that uses X and Z checks for detecting bit flips and phase flips occurring on the physical qubits. These checks stabilize the logical system, so it is also referred to as a stabilizer code \cite{gottesman1997stabilizercodesquantumerror}. Utilizing this a mechanism, a logical qubit can be built by encoding the state over some number of physical qubits. It is well-suited for 2D meshes of qubits, as it works well with the fact that qubits can only interact with their nearest neighbors. A surface code is defined with the notion of $[[n,k,d]]$ where $n$ is the number of physical qubits used, $k$ is the number of maximum encodings that the code can support, and $d$ is the code distance and determines the level of error protection. In this work, it is assumed that $k$ is always one and $d$ is defined by the user. In addition, the structure of the logical qubit within this work is a rotated surface code \cite{PhysRevA.76.012305}. It is illustrated in Figure \ref{fig:surfacecode}. 

\subsection{Lattice Surgery and Logical Gates}
\begin{figure}[!t] % Use !t to force it to the top of the page
    \centering
    
    % --- Row 1: The Wide Image (S and T Gates) ---
    % We keep this container the full width of the text column
    \begin{subfigure}{\columnwidth}
        \centering
        % Kept your original 0.7 scale since this image is naturally wide
        \includegraphics[width=0.7\linewidth]{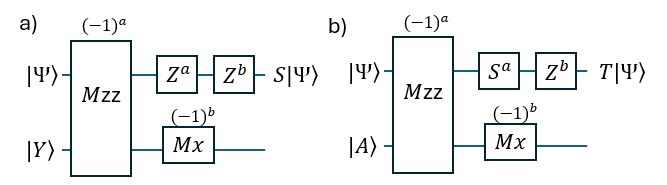}
        \Description{Circuit diagrams for S and T gates using measurement Mzz and corrections.}
        \caption{S and T gates with magic states}
        \label{fig:SandTgates}
    \end{subfigure}
    
    \vspace{1em} % Spacing between the top row and bottom row
    
    % --- Row 2: The Two Smaller Images Side-by-Side ---
    
    % Bottom Left: CNOT
    % We create a container that is roughly half the page width (0.48)
    \begin{subfigure}[b]{0.48\columnwidth}
        \centering
        % Note: We increase width to 1.0\linewidth relative to this SMALLER container
        % This makes it visually roughly the same size as your original 0.5\columnwidth
        \includegraphics[width=\linewidth]{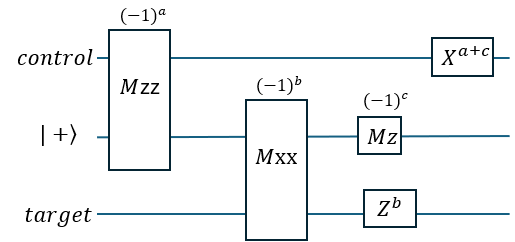}
        \Description{Circuit diagram for a CNOT gate using joint measurements Mzz and Mxx.}
        \caption{CNOT using joint measurements}
        \label{fig:CNOTlattice}
    \end{subfigure}
    \hfill % This is crucial: it pushes the two subfigures apart to the edges
    % Bottom Right: Hadamard
    \begin{subfigure}[b]{0.48\columnwidth}
        \centering
        \includegraphics[width=0.9\linewidth]{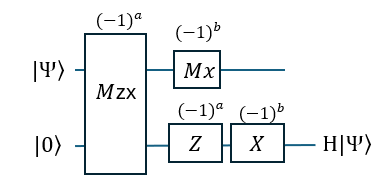}
        \Description{Circuit diagram for a Hadamard gate using measurement Mzx.}
        \caption{Hadamard gate}
        \label{fig:hadamard}
    \end{subfigure}

    % --- Main Caption ---
    \caption{Universal gate set operations that require lattice surgery. System uses (a) S and T gates, (b) CNOT gate, and (c) Hadamard gate. Pauli operations are done via software.}
    \label{fig:all_gates}
\end{figure}

Lattice surgery is a clever method of executing logical gates by using additional qubits between logical patches. It does so without destructing the ability of error detection by turning on and off stabilizers in between patches. It can merge two logical states into one state, or it can split a logical composite state into two logical states. Usually, it is assumed that there are physical ancilla qubits between the logical patches that are either in the ground state or in the plus state. These qubits can be used to create additional stabilizers between two patches. After checking these stabilizers $d$ rounds, where $d$ is the distance of the code, the states of the two logical qubits are merged into a single state. This process is called surface merging. After the merging operation, by measuring the boundary data qubits of these two patches and turning off the stabilizers, the single state can be split into two logical states. This operation is called surface splitting. For more details of LS we refer to \cite{Horsman_2012}. The concept of LS changes some of the fundamentals of quantum gates. For instance, to be able to do a CNOT gate between two logical qubits there is a need for a logical ancilla qubit. Also, the Hadamard gate utilizes a logical ancilla to prevent rotating the code space. In Figure \ref{fig:SandTgates}, S and T gates are illustrated with magic states $|Y\rangle = (|0\rangle + i|1\rangle)/\sqrt{2}$ and $|A\rangle = (|0\rangle + e^{i\pi/4}|1\rangle)/\sqrt{2}$ respectively.

% \begin{figure} 
%     \centering
%     \includegraphics[width=0.7\linewidth]{myfigures/SandTgates.png}
%     \caption{(a) S and (b) T gates with magic states}
%     \label{fig:SandTgates}
% \end{figure}

In Figure \ref{fig:CNOTlattice}, the CNOT gate is illustrated. In total, it uses two joint measurements and, to perform it correctly, one additional ancillary patch is needed. It is assumed that this fault-tolerant CNOT gate is completed in $2d$ cycles where $d$ is the distance of the logical qubit.

% \begin{figure}  
%     \centering
%     \includegraphics[width=0.5\linewidth]{myfigures/CNOT.png}
%     \caption{CNOT using joint measurements}
%     \label{fig:CNOTlattice}
% \end{figure}

In Figure \ref{fig:hadamard}, the Hadamard gate is illustrated. In total, it uses one joint measurement and to perform it without rotating the codespace, one additional ancillary patch is needed. It is assumed that this fault-tolerant Hadamard gate is completed in $d$ cycles.

% \begin{figure}  
%     \centering
%     \includegraphics[width=0.45\linewidth]{myfigures/Hadamard.png}
%     \caption{Hadamard gate}
%     \label{fig:hadamard}
% \end{figure}

\begin{figure} 
    \centering
    \includegraphics[width=0.40\linewidth]{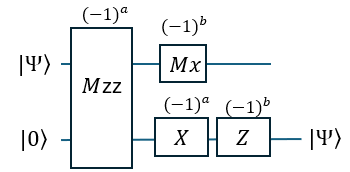}
    \caption{Smooth teleportation protocol of a logical state}
    \label{fig:smooththel}
\end{figure}

In Figure \ref{fig:smooththel}, the Smooth Teleportation protocol is illustrated. This gate enables the teleportation of a logical state from one patch to another by applying a series of joint merge and split measurements and a single measurement of the logical state \cite{Erhard_2021}. The resulting information is then used to apply Pauli corrections on the transferred logical state. Within this framework, the Smooth Teleportation gate is frequently employed to route the desired state from logical data qubit to an ancilla and subsequently across the ancilla network to other ancillas within the same core. For inter-core state transfer, the same gate is utilized; however, when a state reaches the ancilla that is located at the edge of a core, it uses EPR pairs to teleport the state to the other nearest neighboring core. More detailed explanations are given in the following sections. 

Throughout this framework, we utilize  a universal gate set consisting of CNOT, H, T, S, and Pauli gates (X,Y,Z). The correction gates that come from the measurement results are assumed to be done on the software side. Even though the S gate correction within the T gate is in the Clifford gate set, we still include it in our analysis and assume it is done on the qubit since it requires LS and a magic state to be done fault-tolerantly.

\section{Prior work}
From a circuit-compilation and runtime analysis of quantum circuits perspective, there are fundamental approaches to modular quantum architectures. Since cores would be located far away from each other on a real hardware, the compilation of the quantum circuit should place the qubits in these cores so that the cost of having inter-core state transfers is minimized. This problem is NP-hard \cite{10228915}. It is usually considered as balancing subgraphs while minimizing the cuts between these subgraphs, which in our case are quantum cores. Various methods are proposed to achieve better mapping efficiency for DQC. Reducing collective communication by using communication buffers \cite{10.1145/3613424.3614253}, converting this problem into a dynamic network flow \cite{10.1145/3579367} problem, or using Reinforcement Learning agents \cite{promponas2024compilerdistributedquantumcomputing,10992725}. There are other methods that have been developed by taking into account the time slices and creating a cost matrix for non-local operations, then minimizing the total cost by using graph partitioning algorithms \cite{Baker_2020} or assignment algorithms \cite{10.1145/3655029}. 

Overall, our work differs from these previous approaches, as we also consider the routing of magic states as a non-local operation. In a real hardware, it makes sense to have a different unit that specializes to generate and distillate magic states. Additionally, in our work we consider the gate operations for logical qubits both locally and non-locally. This makes our work more complete and close to real-world hardware. Furthermore, by explicitly modeling both local and distributed operations, our approach captures the impact of classical and quantum communication on performance and resource usage \cite{escofet2025impactclassicalquantumcommunication}. This allows for a more realistic evaluation of scheduling, routing, and entanglement management, which are essential for scalable fault-tolerant quantum computing.
We have gathered several concepts together to have a more complete resource analysis of DQC with logical qubits.

\section{Proposed Architecture for DQC}
\begin{figure} 
    \centering
    \includegraphics[width=0.55\linewidth]{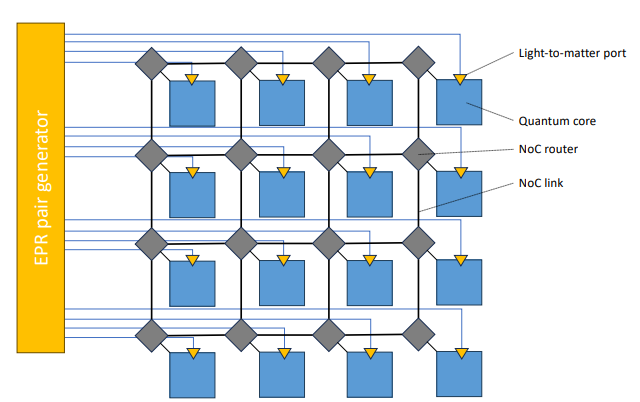}
    \caption{Multi-core quantum architecture \cite{10819534}}
    \label{fig:complete_system}
\end{figure}

A previous work was done \cite{10819534} that assumed an all-to-all connection within a core, with each core directly hosting physical qubits. This structure is illustrated in Figure \ref{fig:complete_system}. In this architecture, all the cores have access to an EPR pair generator and a classical Network on Chip (NoC) of links and routers. By enabling both local and distributed operations, such a modular approach provides a pathway toward scalable quantum computation while mitigating limitations arising from hardware such as noise, connectivity, and dense control electronics. 

However, this proposal does not account for logical qubits, the routing of magic states needed for logical gates, logical qubit operations via LS, and architecture optimization for mapping the quantum circuits to such structures.

In this regard, we propose an extension to such architectures that includes a 2D mesh of quantum cores each hosting an equal number of logical data qubits and logical ancilla patches. This mesh has a  connection to an EPR pair distribution unit as in Figure \ref{fig:complete_system}. In addition, it has a connection to a MSF as in Figure \ref{fig:network}. Using these two blocks, which act as support units to the computing unit (which is the distributed cores), our proposal represents a complete system for fault-tolerant DQC. 

\begin{figure}   
    \centering
    \includegraphics[width=0.6\linewidth]{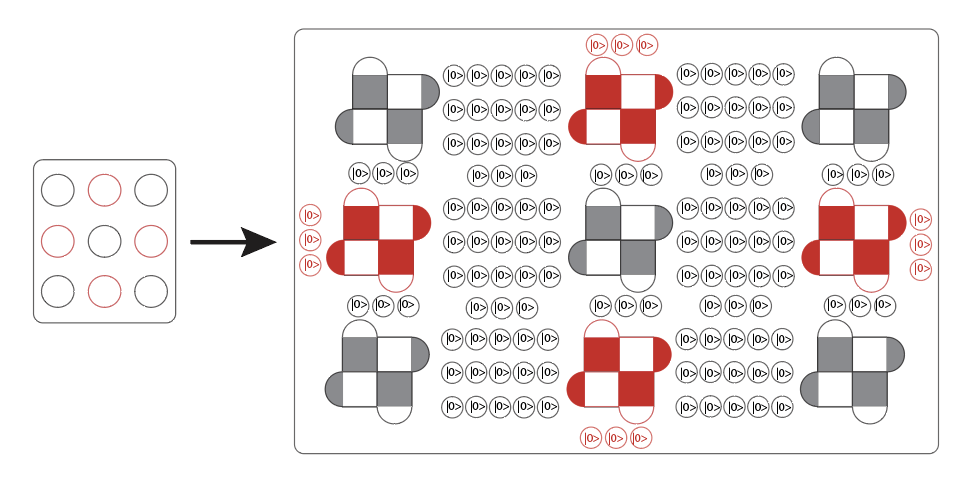}
    \caption{Structure of one core}
    \label{fig:structureofonecore}
\end{figure}

In Figure \ref{fig:structureofonecore}, the inside of a core is shown. The red rotated surface code patches are logical ancillas, and the gray patches are logical data qubits. The remaining qubits in between patches are used to implement merge and split operations for LS. The physical qubits on the core edges (red circles) are assumed to be used for inter-core operations. 

\begin{figure}
    \centering
    \includegraphics[width=0.7\linewidth]{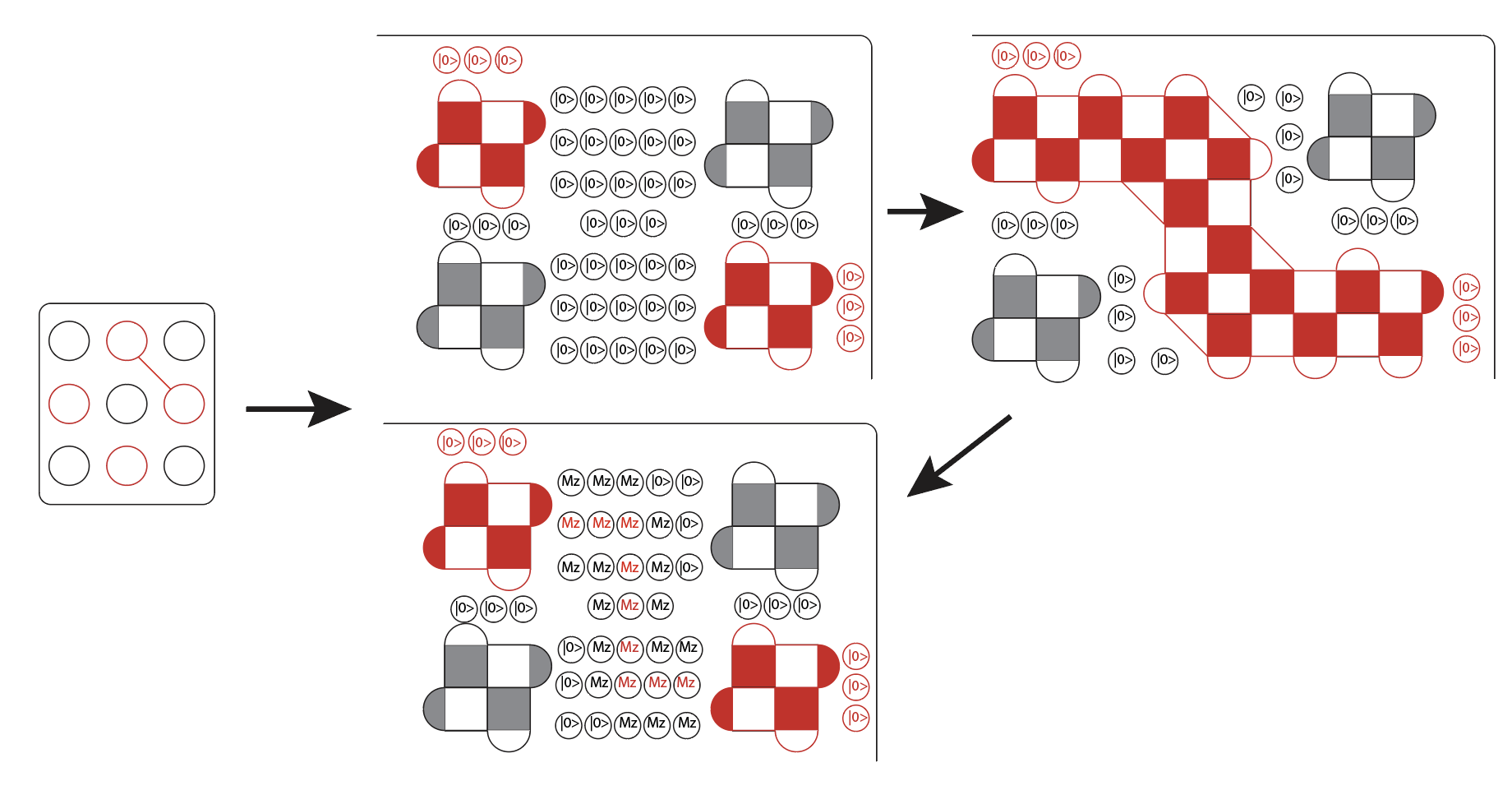}
    \caption{Diagonal Merge and Split}
    \label{fig:diagonalmergeandsplit}
\end{figure}

In this framework, it is assumed that the diagonal merge and split operations can be performed using the idea in \cite{fowler2019lowoverheadquantumcomputation} by utilizing the qubits in between two nearest logical ancilla to satisfy the merge and split operations. This is illustrated in Figure \ref{fig:diagonalmergeandsplit}. This feature forms the basis of routing states over the ancilla network for doing distant CNOTs and magic state travel. In this setup, whenever a logical qubit needs to do a single-gate operation that requires a magic state, it is routed from the MSF to the nearest ancilla by implementing a series of smooth teleportation and inter-core teleportation. Then, the logical data qubit and the ancilla interact using joint measurements with LS.

\begin{figure} 
    \centering
    \includegraphics[width=0.6\linewidth]{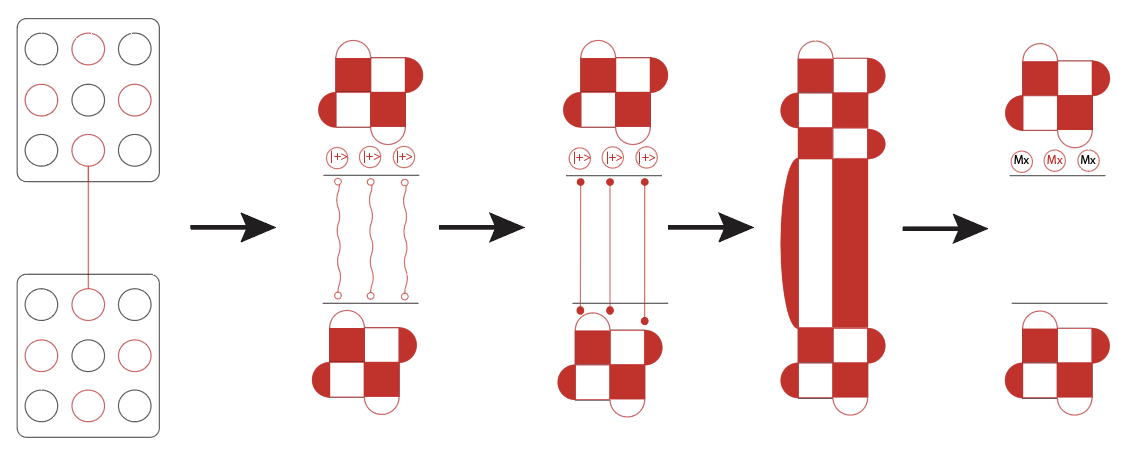}
    \caption{LS between cores}
    \label{fig:coretocorelatticesurgery}
\end{figure}

These cores are connected to each other by facilitating  EPR pair distribution that enables smooth teleportation operations between two distant ancilla patches located in neighboring cores. It is assumed that by using these EPR pairs, the merge and split operations can be done by teleporting the CNOT gate over the cores as in \cite{Ramette2024}. Since every ancilla patch on the core edge has additional physical qubits, CNOTs needed for stabilizer checks can be transferred between cores as in Figure \ref{fig:coretocorelatticesurgery}. The necessary number of EPR pairs for this operation to be fault-tolerant is $O(d^2)$, where $d$ is the code distance. This is because they are consumed every inter-core operation, and stable operations should be checked at least $d$ rounds consecutively. In this framework, it is assumed that EPR pairs can be distributed to each edge logical ancilla on a core. A core with $n^2$ logical qubits has $(n^2+1)/2$ logical data qubits and $(n^2-1)/2$ logical ancilla qubits. In total, a core has $(n-1)/2$ logical ancillas per edge and $2n-2$ total logical ancillas that can access EPR pairs to transfer a state to the nearest core.

On a real hardware, this configuration makes sense because the communication would be bottlenecked if the cores had very few quantum channels. In such a case, the scheduler would be unable to efficiently parallelize the gates, as there would not be space for different state transfers at the same time. Building such a structure is also necessary for the proposed idea of utilizing the ancilla network for transferring magic states and enabling distant CNOTs.

\section{Fault Tolerant Operations}
The first part of simulating a quantum circuit on our architecture requires decomposing the circuit to a set of instructions that tell the system what to execute during each time slice. While systems using physical qubits just need to communicate locations of physical qubits and their corresponding gates, the multi-step structure of fault-tolerant operations on a logical qubit requires a system for scheduling and routing. We also take measures in employing path-finding methods in our architecture to pipeline the execution of fault-tolerant gates. 

\subsection{Order of Operations}
\begin{figure}
    \centering
    \includegraphics[width=0.5\linewidth]{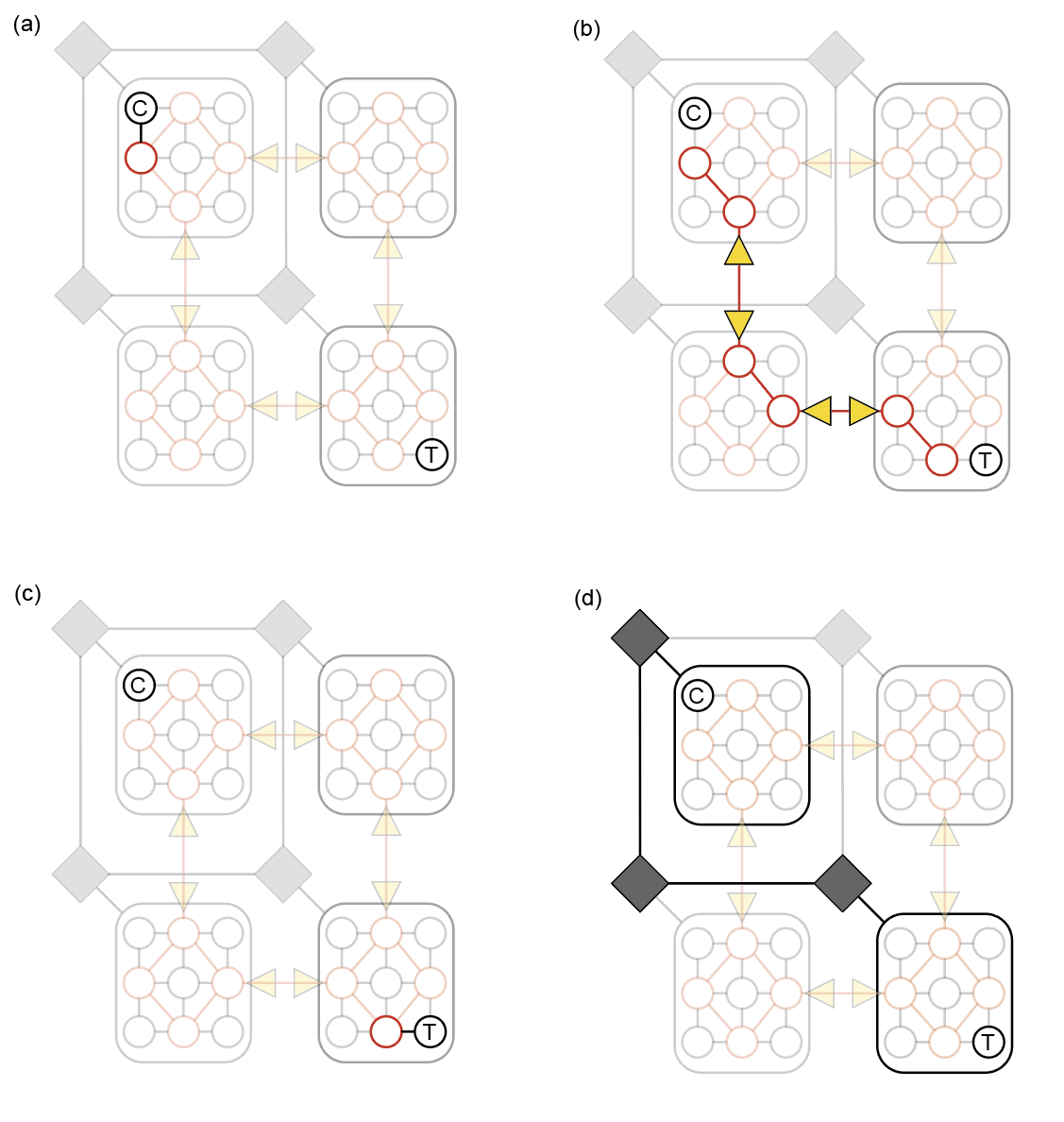}
    \caption{Execution of a CNOT gate by (a) instantaneous transversal preparation of adjacent logical ancilla into the plus state and Mzz operation between logical ancilla and control logical data qubit, (b) movement of the state of logical ancilla, (c) Mxx operation between logical ancilla and target logical data qubit, and (d) classical communication of results between cores on NoC}
    \label{fig:CNOT}
\end{figure}

Supporting LS operations within the architecture requires defining the sequencing of both resource allocation and LS. Single-qubit gates require logical ancilla to be prepared in order to perform a LS operation. The Hadamard and Smooth Teleportation gates require logical ancilla to be initialized in either the plus or zero state, which can be done instantaneously on available on-core ancilla. In contrast, the S and T gates require magic states that must first be distilled in an MSF off the mesh and then transferred to an available ancilla on the core. The two-qubit CNOT gate also requires an ancilla prepared in the plus state; however, the structure of the fault-tolerant operation can be leveraged so that logical qubits remain stationary, while the ancilla state is transferred. Figure \ref{fig:CNOT} shows the stationary implementation of the CNOT in our architecture. This differs from current physical qubit distributed works, where all participating qubits must reside within the same core to perform multi-qubit gates. By exploiting the sequencing of LS operations, we have developed an innovative sequence and routing structure that keeps data logical qubits fixed in place, enabling new optimization strategies for quantum processing units within a mesh.

\subsection{Scheduling and Routing Operations}
As discussed, logical data qubits within our architecture remain fixed while ancilla logical qubits are used to support fault-tolerant gate operations. At any moment within the circuit, the ancilla are either unused or being used within an LS operation. We introduce a network of ancilla within our architecture in order to schedule and route operations.

\begin{figure}
    \centering
    \includegraphics[width=0.5\linewidth]{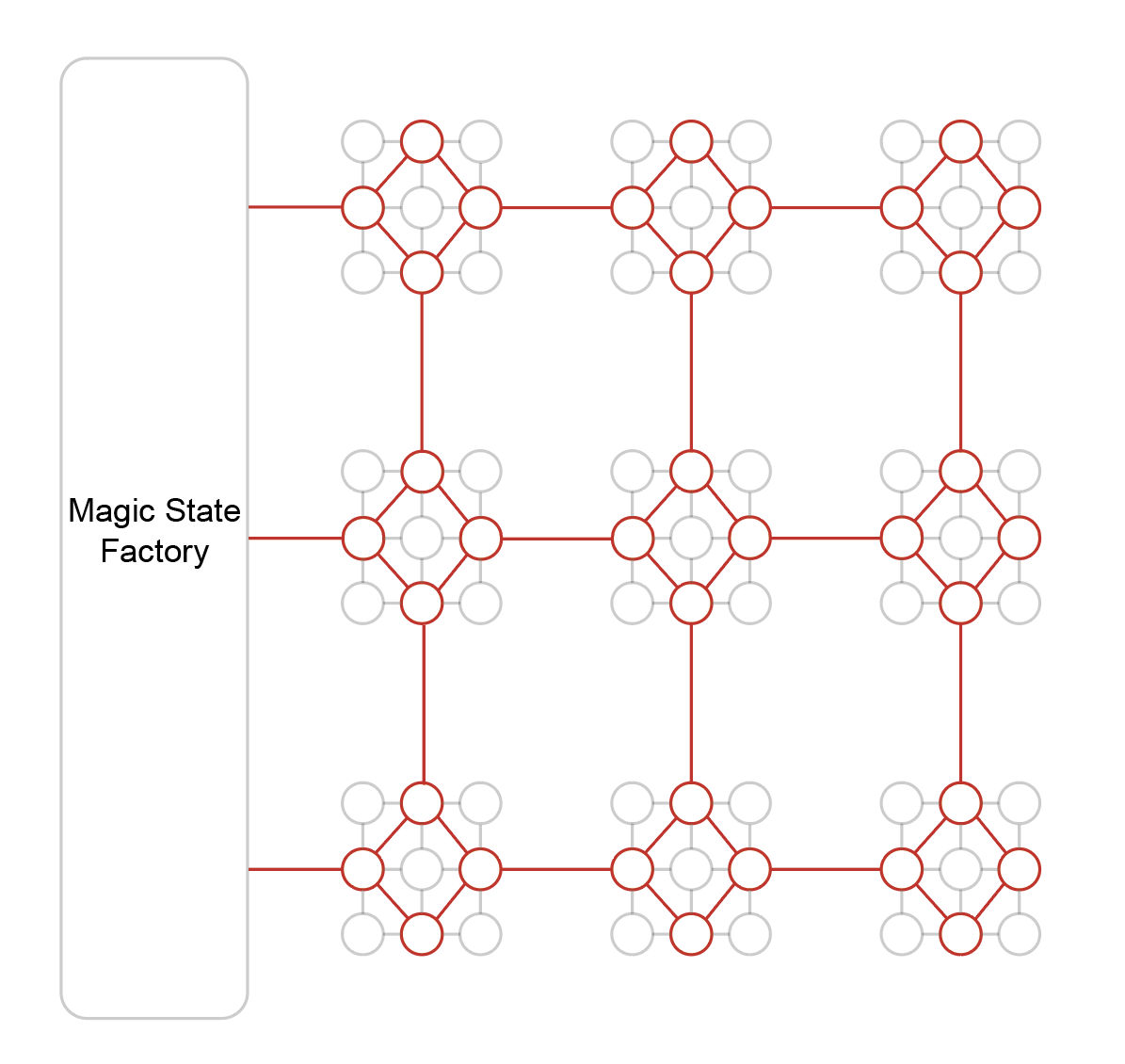}
    \caption{Ancilla network can be represented as a graph for path finding methods}
    \label{fig:network}
\end{figure}

\begin{table}
\caption{Timing and logical ancilla allocation used within the scheduler for fault tolerant gates where $d$ is the code distance.}
\centering
\small % make text slightly smaller to fit
\renewcommand{\arraystretch}{1.3}
\setlength{\tabcolsep}{4pt} % reduce horizontal padding
\begin{tabularx}{\columnwidth}{@{}p{3.5cm} >{\centering\arraybackslash}m{7cm} @{\hspace{8pt}} X@{}}
\toprule
\textbf{Gate Type} & \textbf{Cycles} & \textbf{Logical Ancilla State} \\
\midrule
Pauli & 0 & None \\
\midrule
S & $d$ & Magic State \\
\midrule
T & $d/2d$ & Magic State \\
\midrule
Hadamard & $d$ & Transversal Zero State \\
\midrule
Smooth Teleportation & $d$ & Transversal Plus State \\
\midrule
Controlled-NOT & $2d$ & Transversal Plus State \\
\bottomrule
\end{tabularx}
\label{tab:timing}
\end{table}

We can represent our network of ancilla as a weighted graph as shown in Figure \ref{fig:network}. The vertices are the ancilla themselves and the edges represent potential merge operations between different ancilla logical qubits. Scheduling the gate operations now devolves into a path-finding exercise considering the order of operations in a fault-tolerant gate. We implement a modified Dijkstra's algorithm (Algorithm \ref{alg:schedule}) within our work to handle this path finding. Dijkstra’s algorithm is a search algorithm used to find the shortest path between nodes in a weighted graph with non-negative edge weights \cite{Dijkstra_1959}. Within our algorithm, the weight of an edge between two vertices is given by the larger execution end time plus the time needed for state movement. This is done in order to minimize the total execution time of the circuit. The time required for each operation is reported in Table \ref{tab:timing}. However, the weights of an edge can be changed depending on what is being optimized. For example, if we only want to minimize total core-to-core operations, we can set all inner core edges to an equal weight, all core-to-core edges to a higher weight, and then sequence parallel operations. 

\begin{algorithm}[!t]
\caption{Scheduling fault-tolerant operations}
\label{alg:schedule}
\begin{algorithmic}[1]
\Function{Schedule}{$CIRCUIT$}
  \ForAll{$GATE \in CIRCUIT$}
    \If{$GATE$ is $CNOT$}
        \State $PATH \gets \textsc{Dijkstra}(q_1, q_2)$
        \State \textbf{map} $\left(\mathrm{{CNOT}_{MZZ}}, \mathrm{PATH}\right) \rightarrow \left(q_1, {ancillas}\right)$
        \State \textbf{map} $\left(\mathrm{PATH}, \mathrm{{CNOT}_{MXX}}\right) \rightarrow \left(q_2,{ancillas}\right)$
    \ElsIf{$GATE$ is $T$ or $S$}
        \State $PATH \gets \textsc{Dijkstra}(factory, q_1)$
        \State \textbf{map} $(PATH, GATE) \rightarrow (q_1, ancillas)$
    \ElsIf{$GATE$ is $H$}
        \State \textbf{map} $(GATE) \rightarrow (q_1, ancillas)$
    \EndIf
  \EndFor
\EndFunction
\end{algorithmic}
\end{algorithm}

\begin{figure}
    \centering
    \includegraphics[width=1\textwidth]{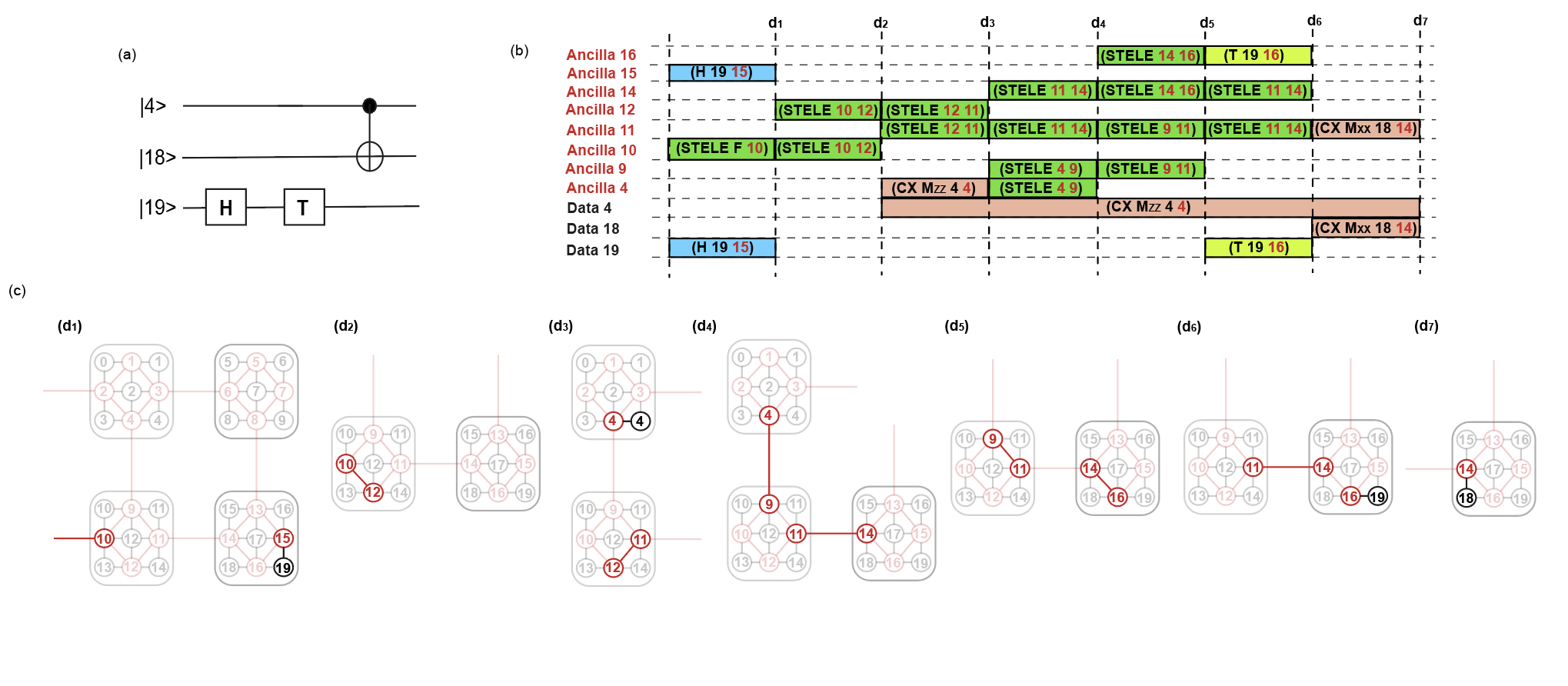}
    \caption{(a) Sample circuit parsed to the (b) scheduler with (c) execution of the circuit at each cycle}
    \label{fig:wide-figu-=re}
\end{figure}

Within the simulation our path finding algorithm tests all available adjacent ancilla, and adjacent ancilla combinations in the case of two qubit gates, and chooses whichever produces the smallest increase in execution time. The time steps within our algorithm are given by the cycles of syndrome measurement, which depends on the code distance. The scheduler also takes steps to ensure the accuracy of the circuit. We prevent idling of logical qubits by delaying the start of CNOT gates if the best path of the ancilla starts at some time after the end of the joint Z operation. Furthermore, even though the joint Z for the CNOT operation only takes $d$ cycles, the end time of the execution in the scheduler is set to the end time of the joint X operation so that no operations can occur on the logical data qubit until post-processing has occurred. Both of these concepts are shown in the example scheduler in Figure \ref{fig:wide-figu-=re}. The scheduler naturally routes and parallelizes the execution of the quantum circuit while not interfering with the circuit's logic. It acts agnostic to other systems in the layout, such as the NoC and EPR pair communication, but does account for the time needed to distill magic states.

\begin{figure}
    \centering
    \includegraphics[width=0.25\linewidth]{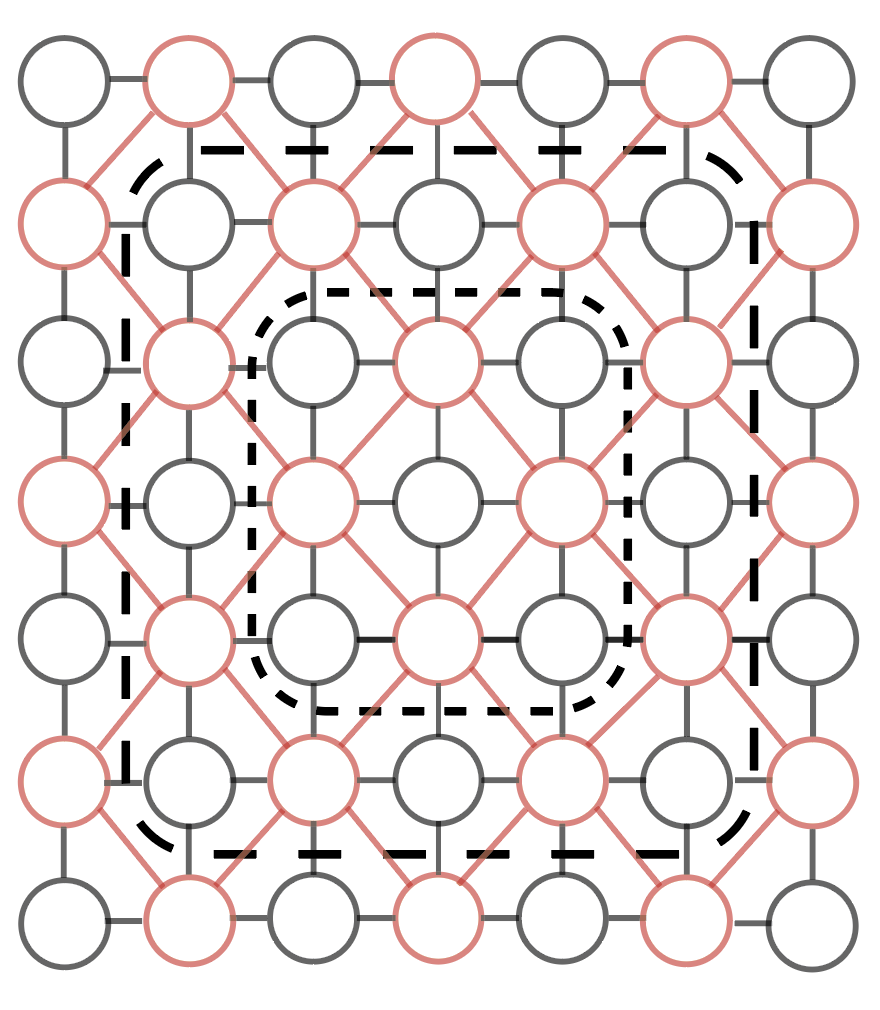}
    \caption{Symmetric scaling of core with logical qubit placement at sizes 3, 5, and 7}
    \label{fig:corescale}
\end{figure}

For Dijkstra's algorithm \cite{Dijkstra_1959} implemented with a binary heap, the complexity is $O((|E| + |V|) \log |V|)$. Within our graph, the nodes are logical qubits, and the edges are potential merge operations between logical qubits facilitated by physical qubits. The parameters of a graph with symmetric cores are defined as modular dimensions $m_x$ and $m_y$, core dimension $c$, and number of gates $N$. The number of qubits within our system, $Q$, can be given as $m_xm_yc^2$. The number of connections between adjacent cores and the magic state factory is $(m_xm_y+m_x(m_y-1))((c-1)/2)$. The first term of the unscaled equation represents the horizontal connections, which include connections to the MSF, and the second term represents the vertical connections. The connections within a core is given by $2c(c-1)+(c-1)^2$, where the first term represents non-diagonal connections, and the second term represents diagonal connections. Note that  the number of edges in a core scales by a  $c^2$ factor due to the scaling of cores as seen in Figure \ref{fig:corescale}. The edges within the core must also be scaled by the number of cores within a system, which is $m_xm_y$. It can then be seen that the total number of edges within the graph is asymptotically bounded by the $m_xm_yc^2$ term, which is equivalent to the number of logical qubits within the system. The complexity of the algorithm simplifies to $O(Q \log Q)$. The scheduling algorithm iterates over the set of gates, so the final complexity can be given by $O(N \cdot Q\log Q)$.

\section{Systems Optimization}
While our scheduler takes steps to pipeline instructions, further operations can be taken to optimize the time and resources used in the circuit execution. The architecture has two layers that we jointly optimize. We approach both optimizations by trying to limit the amount of inter-core operations (Algorithm \ref{alg:optimize}). Two operations lead to inter-core movement, movement of magic states for a T or S gate and movement of an ancilla state for multi-qubit operations. The first relates to how qubits are assigned to a core and where cores are placed in relation to each other on the mesh. Meanwhile, the optimization of magic state routing relates to where cores are placed in relation to the distillation factory. Within both parts of the optimization, we leverage the fact that logical data qubits within our architecture are stationary.
\subsection{Qubits to Core Optimization}
\begin{figure}
    \centering
    \includegraphics[width=0.6\linewidth]{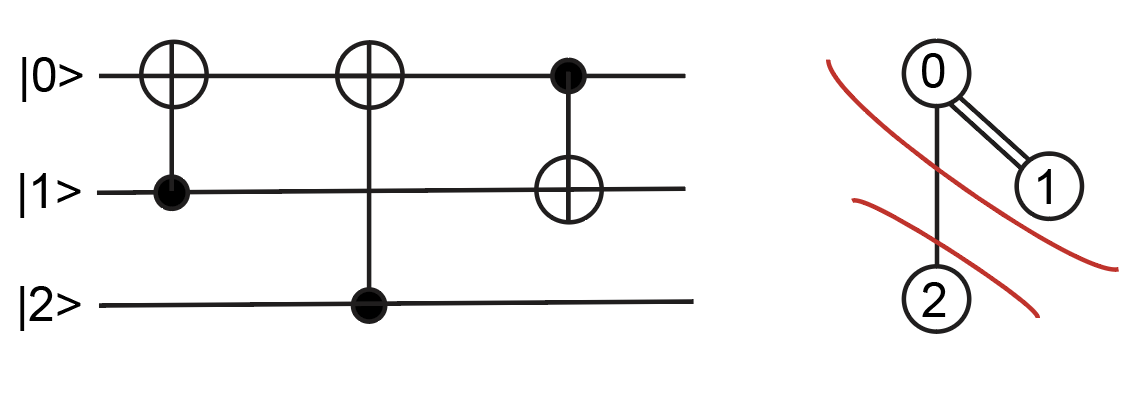}
    \caption{Example of partitioning a circuit between two cores}
    \label{fig:placeholder}
\end{figure}

Within the first part of the algorithm, we utilize a graph partitioning method to minimize the amount of inter-core CNOT operations. As seen in Figure 
\ref{fig:placeholder}, the vertices represent the qubits and the edges represent a CNOT operation. In partitioning this graph, we group the qubits into equal sized cores, minimizing the amount of CNOT operations that require movement between cores. We use KaHIP partitioning within our simulation to handle the zero imbalance partitions \cite{DBLP:conf/wea/SandersS13}.

\subsection{Optimizing Core Placement Within a Mesh}
\begin{figure}
    \centering
    \includegraphics[width=0.6\linewidth]{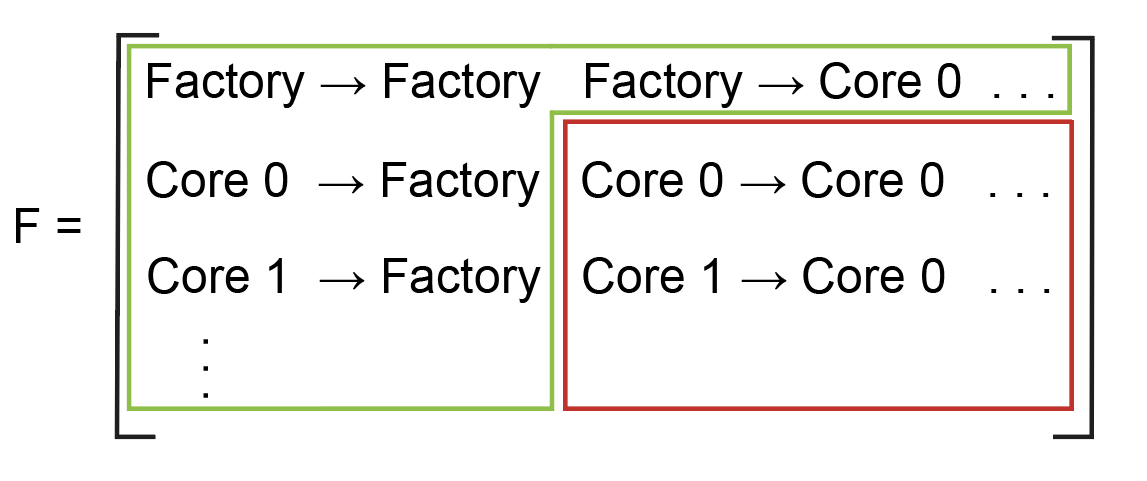}
    \caption{Flow matrix used in quadratic assignment where the red box represents inter-core operations from CNOTS collected by the edges between partitons and the green box represents the amount of inter-core operations from magic state routing }
    \label{fig:flowmatrix}
\end{figure}

\begin{figure}
    \centering
    \includegraphics[width=0.6\linewidth]{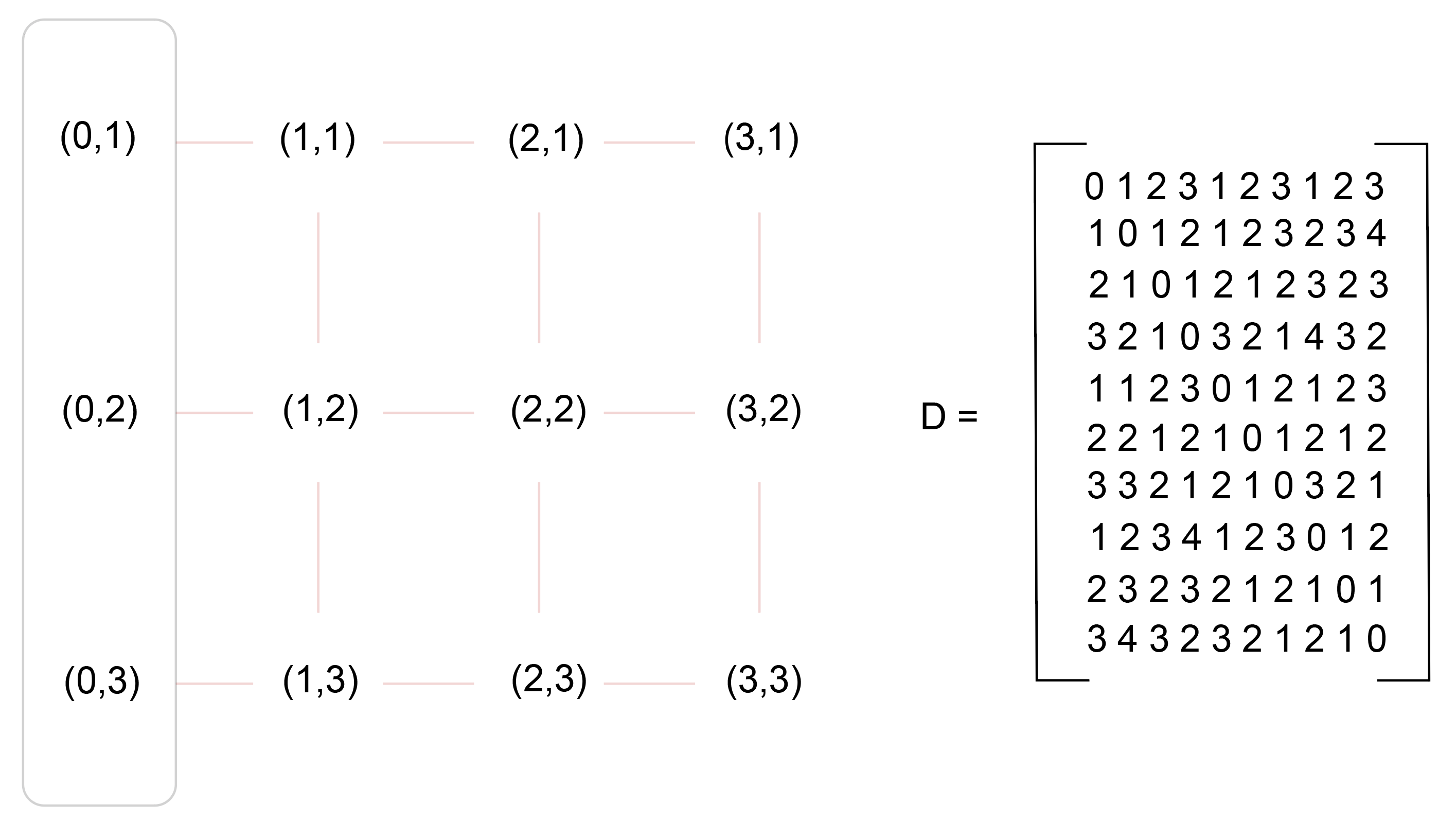}
    \caption{Distance matrix for a 6x6 mesh}
    \label{fig:distancemat}
\end{figure}
After partitioning the qubits we are left with a set of cores and the leftover edges between them. Furthermore, we still have not considered the inter-core operations from routing magic states. Both of these criteria are handled when optimizing the placement of cores onto the mesh. This is a novel approach, as past optimization models usually consider an all-to-all connection, neglecting movement of states along the mesh and from a system (MSF) off the mesh. In the second part of our optimization algorithm, we have two variables to consider, the distance between locations on the mesh and the amount of core-to-core and factory-to-core inter-core operations. By rephrasing inter-core operations as the amount of flow between cores and prep states, the problem reduces to minimizing the total amount of flow. 

The optimization problem becomes a quadratic assignment problem (QAP), as introduced in \cite{6b66faf6-3f33-3aa9-8609-22b8cfa3f37a}. QAP is finding an optimal assignment of facilities to locations in order to minimize the total cost, where the cost depends on both the flow between facilities and the distance between locations. This problem is known to be NP-hard, and even approximate solutions become computationally challenging as the system size increases. 

We want to assign cores to the mesh in order to minimize the amount of inter-core operations. The problem is:

\begin{equation}
\text{minimize      } \sum_{i,j\in[n]} f_{i,j}d_{\pi(i)\pi(j)} \text{      where } \pi \in \prod
\end{equation}

    $f_{i,j}$ represents the flow between the cores $i$ and $j$, the amount of inter-core operations, and $d_{\pi(i)\pi(j)}$ represents the distance between locations in the mesh, assigned by the current permutation of $i$ and $j$. In a mesh topology, the distance is given by the Manhattan distance. The flow matrix structure that uses the results of the graph partitioning can be seen in Figure \ref{fig:flowmatrix}, and an example distance matrix can be seen in Figure \ref{fig:distancemat}. We can account for the prep state module by considering $N+1$ cores and fixing the location of the first spot in all permutations, so we have permutations of size $N+1$ but only permutating over $N$ values. This will allow us to consider the flow of both inter-core operations from CNOTs and inter-core operations from routing the magic states. The problem then becomes a quadratic assignment problem with fixed assignment (QAPFA). The problem is NP-hard, but it can be solved exactly for fewer than ten cores in a reasonable amount of time. For larger meshes, we implement the heuristic Robust Taboo Search as it can efficiently handle larger sets of cores and the iterative approach makes it an ideal choice for implementing a fixed permutation \cite{Taillard_1991, Taillard_1995}.

While we use leftover edges from the partitions to construct the core-to-core inter-core operations, we use a weighted system for assigning inter-core operations between the factory and cores. The S gates are weighted to a constant since they require one magic state, while the T gates are weighted as 1.5 $\times$ the constant since we give it a probability of also requiring an S gate following the execution of a T gate. The overall flow of the algorithm involves first partitioning the cores, then using the partitioned graph to construct the flow matrix while the distance matrix is constructed from the inputted mesh, and lastly either directly solving the quadratic assignment problem or using our chosen heuristic.

\begin{algorithm}
\caption{Optimizing mapping of logical cores and qubits}
\label{alg:optimize}
\begin{algorithmic}[1]
\Function{optimize}{$CIRCUIT$}
  \ForAll{$GATE \in CIRCUIT$}
    \If{$GATE$ is $CNOT$}
        \State $\textsc{create\_edge}(GRAPH,q_1, q_2)$
    \ElsIf{$GATE$ is $S$}
        \State $WEIGHT[q_1] \mathrel{+}= weight\_constant$ 
    \ElsIf{$GATE$ is $T$}
        \State $WEIGHT[q_1] \mathrel{+}= weight\_constant \cdot 1.5$ 
    \EndIf
  \EndFor
  \State $CORES \gets \textsc{partition}(GRAPH)$ 
  \ForAll{$CORE \in CORES$}
    \State $FLOW[\mathrm{CORE}_i][\mathrm{CORE}_j] \gets$
    \State \hspace{1em} $\textsc{count\_edges}(\mathrm{CORE}_i, \mathrm{CORE}_j)$
  \EndFor
  \ForAll{$LOC \in MESH$}
    \State $DIST[\mathrm{LOC}_i][\mathrm{LOJ}_j] \gets$
    \State \hspace{1em} $\left| x_i - x_j \right| + \left| y_i - y_j \right|$
  \EndFor
  \For{$i = 0$ \textbf{to} $NUM\_CORES$}
     \State $\mathrm{CORE}_i \gets CORES[i]$ 
    \State $FLOW[\mathrm{FACTORY}][\mathrm{CORE}_i] \gets$ 
    \State \hspace{1em} $SUM(WEIGHT[qubit] \textbf{ for } \mathrm{ qubit } \textbf{ in }  \mathrm{CORE}_i)$\
    \State $FLOW[\mathrm{CORE}_i][\mathrm{FACTORY}] \gets$
    \State \hspace{1em} $FLOW[\mathrm{FACTORY}][\mathrm{CORE}_i]$ 
    \State $DIST[\mathrm{FACTORY}][\mathrm{CORE}_i] \gets$ $x_i$
    \State $DIST[\mathrm{CORE}_i][\mathrm{FACTORY}] \gets$ 
    \State \hspace{1em} $DIST[\mathrm{FACTORY}][\mathrm{CORE}_i]$
  \EndFor
  \State $MAP \gets$ \textsc{quadratic}(FLOW, DIST)
\EndFunction
\end{algorithmic}
\end{algorithm}

\subsection{Qubit Placement Within a Core}
This part of the optimization assigns qubits to a core by sorting each core’s qubit list that comes from the previous step according to a precomputed weight vector, where each weight reflects the total number of T and S gates associated with that qubit. By ordering the qubits in descending weight, the most resource-intensive qubits in regard of magic states are placed closest to the MSF. The remaining qubits follow in decreasing priority, ensuring that placement aligns with computational demand (Algorithm \ref{alg:insidecoreplace}). This would conclude the hierarchical optimization within our model regarding the mapping of the qubits.
\begin{algorithm}
\caption{Optimizing mapping of qubits in a core}
\label{alg:insidecoreplace}
\begin{algorithmic}[1]
\Function{InsideCorePlacement}{$MAPPEDCORES$,$WEIGHT$}
  \For{$i = 0$ \textbf{to} $NUM\_CORES-1$}
     \State \textbf{sort} $\mathrm MAPPEDCORES[i]$
     \State \textbf{for} $\mathrm {a} \textbf{ and } \mathrm{b} \textbf{ in } MAPPEDCORES[i]$
     \State \textbf{respect to} $WEIGHT[a] > WEIGHT[b]$
  \EndFor
  \State $MAP \gets$ \textsc{Fill first close to MSF}
\EndFunction
\end{algorithmic}
\end{algorithm}
\section{Results}
The proposed framework was tested with different quantum circuits, generated from Qiskit \cite{gadi_aleksandrowicz_2019_2562111}, and different layouts to demonstrate the feasibility of optimizing core placement in a 2D mesh. First, only a random placement of qubits to cores is implemented. Then, only the KaHIP algorithm was used to place the qubits more efficiently to the cores to reduce the inter-core CNOTs. Lastly, the new optimization stage that is introduced in this work was implemented together with KaHIP. The results clearly indicate that even for small number of gates there is a significant improvement on the number of executed gates, EPR pairs, total travel of magic states, and total cycles of the circuit. Different circuits with varying core and mesh sizes, gate probabilities, and number of gates were analyzed to find the best structures and see the best gains in terms of statistics. This is essential to reduce the undesired cost of these modular architectures and find the most efficient layout.

Table \ref{table:2} shows an improvement in all statistics after optimizing the core placement in the mesh. The circuit used for this benchmark has a six-to-four ratio of one- and two-qubit gates. It is shown that by using the new optimization on the mapping side, the total travel of the magic states was significantly reduced. It also reduced the total number of gates, EPR pairs, and total cycles.

\begin{table}
\caption{Random 10,000 Gate Circuit | 6x6 Mesh | 5x5 Core Size | Code Distance 3 | 468 Logical Qubits | 60/40 one-qubit gate/two-qubit gate.}
\centering
\small % make text slightly smaller to fit
\renewcommand{\arraystretch}{1.3}
\setlength{\tabcolsep}{4pt} % reduce horizontal padding
\begin{tabularx}{\columnwidth}{@{}p{4cm} >{\centering\arraybackslash}m{3cm} >{\centering\arraybackslash}X >{\centering\arraybackslash}X@{}}
\toprule
\textbf{Stats-Mapping} & \textbf{Part.+ QAPFA} & \textbf{Part.} & \textbf{Random} \\
\midrule
Executed Gates & 110448 & 114100 & 129257 \\
\midrule
EPR Pairs & 241623 & 250785 & 288090 \\
\midrule
Magic State Travel & 40868 & 43191 & 45088 \\
\midrule
Cycles & 10944 & 11298 & 13701 \\
\bottomrule
\end{tabularx}
\label{table:2}
\end{table}

Table \ref{table:3} shows a circuit with an uneven distribution of gates. Half of the logical qubits have significantly more CNOTs than T and S gates while the other half has the reversed probability. Our placement algorithm shows a strong improvement on the total travel of magic states. This is due to our optimization algorithm placing cores with a high number of gates that utilize magic states closer to the MSF. This benchmark is a good example of the benefits of using our algorithm.

\begin{table}
\caption{Random 10,000 Gate Circuit | 6x6 Mesh | 5x5 Core Size | Code Distance 3 | 468 Logical Qubits | Clustered 80/20 20/80 one-qubit gate/two-qubit gate.}
\centering
\small % make text slightly smaller to fit
\renewcommand{\arraystretch}{1.3}
\setlength{\tabcolsep}{4pt} % reduce horizontal padding
\begin{tabularx}{\columnwidth}{@{}p{4cm} >{\centering\arraybackslash}m{3cm} >{\centering\arraybackslash}X >{\centering\arraybackslash}X@{}}
\toprule
\textbf{Stats-Mapping} & \textbf{Part.+ QAPFA} & \textbf{Part.} & \textbf{Random} \\
\midrule
Executed Gates & 93058 & 106584 & 136924 \\ 
\midrule
EPR Pairs & 196308 & 229536 & 302337 \\
\midrule
Magic State Travel & 22885 & 35402 & 35751  \\
\midrule
Cycles & 11145 & 11187 & 15390  \\
\bottomrule
\end{tabularx}
\label{table:3}
\end{table}

Table \ref{table:4} has a circuit with the same probability for all the gates. It has a significant gain in all statistics. Although the number of gates, number of qubits, and the layout is the same as previous benchmarks, it results in lower executed gates and EPR pairs consumed since it has a lower amount of CNOTs.

\begin{table}
\caption{Random 10,000 Gate Circuit | 6x6 Mesh | 5x5 Core Size | Code Distance 3 | 468 Logical Qubits | Same Probability for All the Gates.}
\centering
\small % make text slightly smaller to fit
\renewcommand{\arraystretch}{1.3}
\setlength{\tabcolsep}{4pt} % reduce horizontal padding
\begin{tabularx}{\columnwidth}{@{}p{4cm} >{\centering\arraybackslash}m{3cm} >{\centering\arraybackslash}X >{\centering\arraybackslash}X@{}}
\toprule
\textbf{Stats-Mapping} & \textbf{Part.+ QAPFA} & \textbf{Part.} & \textbf{Random} \\
\midrule
Executed Gates & 79736 & 84326 & 93857 \\ 
\midrule
EPR Pairs & 177615 & 18669 & 212517 \\
\midrule
Magic State Travel & 51768 & 55701 & 56058  \\
\midrule
Cycles & 10470 & 10536 & 13920  \\
\bottomrule
\end{tabularx}
\label{table:4}
\end{table}

Table \ref{table:5} is an example of exactly solving the placement of cores in our mesh instead of using a heuristic. Since the number of cores is just nine, it is well in the upper bound of the size of a mesh for QAPFA to have an exact solution. However, in this benchmark, we kept the number of gates higher than in the previous benchmarks. The results clearly present gains in all statistics when our optimization is used after partitioning. 

\begin{table}
\caption{Random 30,000 Gate Circuit | 3x3 Mesh | 5x5 Core Size | Code Distance 3 | 117 Logical Qubits | 60/40 one-qubit gate/two-qubit gate.}
\centering
\small % make text slightly smaller to fit
\renewcommand{\arraystretch}{1.3}
\setlength{\tabcolsep}{4pt} % reduce horizontal padding
\begin{tabularx}{\columnwidth}{@{}p{4cm} >{\centering\arraybackslash}m{3cm} >{\centering\arraybackslash}X >{\centering\arraybackslash}X@{}}
\toprule
\textbf{Stats-Mapping} & \textbf{Part.+ QAPFA} & \textbf{Part.} & \textbf{Random} \\
\midrule
Executed Gates & 188925 & 190407 & 194393 \\ 
\midrule
EPR Pairs & 383409 & 386181 & 398007 \\
\midrule
Magic State Travel & 55176 & 56515 & 56429  \\
\midrule
Cycles & 50226 & 50646 & 50988  \\[1ex] 
\bottomrule
\end{tabularx}
\label{table:5}
\end{table}

Table \ref{table:6} shows how much the layout of the mesh affects the efficiency of resource consumption. Compared to Table \ref{table:2}, this benchmark has more cores, but each core has a smaller number of qubits. Also, this layout is potentially inefficient since it is not a square and the ratio of cores that can reach the MSF is lower compared to Table \ref{table:2}. Even though the number of qubits is very close, and the number of gates is equal, this structure results in almost more than two times the amount of EPR pairs consumed and more than three times the amount of cycles needed to execute the circuit.  

\begin{table}
\caption{Random 10,000 Gate Circuit | 8x12 Mesh | 3x3 Core Size | Code Distance 3 | 480 Logical Qubits | 60/40 one-qubit gate/two-qubit gate.}
\centering
\small % make text slightly smaller to fit
\renewcommand{\arraystretch}{1.3}
\setlength{\tabcolsep}{4pt} % reduce horizontal padding
\begin{tabularx}{\columnwidth}{@{}p{4cm} >{\centering\arraybackslash}m{3cm} >{\centering\arraybackslash}X >{\centering\arraybackslash}X@{}}
\toprule
\textbf{Stats-Mapping} & \textbf{Part.+ QAPFA} & \textbf{Part.} & \textbf{Random} \\
\midrule
Executed Gates & 118458 & 125823 & 143877 \\ 
\midrule
EPR Pairs & 403767 & 429480 & 500994 \\
\midrule
Magic State Travel & 46702 & 51315 & 52880  \\
\midrule
Cycles & 14007 & 15300 & 18378  \\
\bottomrule
\end{tabularx}
\label{table:6}
\end{table}

Table \ref{table:7} shows the same number of qubits with the same circuit that was used for Table \ref{table:6}, but with a much more efficient inverse configuration. This benchmark has more cores that are connected to the MSF. This enables the flow of magic states from more channels and reduces the traffic of the total flow. Even though the number of qubits is the same and the number of gates is equal, this structure consumes lower EPR pairs, finishes in lower cycles, results in lower total travel of magic states, and the QAPFA optimization gives more gain compared to just partitioning.

\begin{table}
\caption{Random 10,000 Gate Circuit | 12x8 Mesh | 3x3 Core Size | Code Distance 3 | 480 Logical Qubits | 60/40 one-qubit gate/two-qubit gate.}
\centering
\small % make text slightly smaller to fit
\renewcommand{\arraystretch}{1.3}
\setlength{\tabcolsep}{4pt} % reduce horizontal padding
\begin{tabularx}{\columnwidth}{@{}p{4cm} >{\centering\arraybackslash}m{3cm} >{\centering\arraybackslash}X >{\centering\arraybackslash}X@{}}
\toprule
\textbf{Stats-Mapping} & \textbf{Part.+ QAPFA} & \textbf{Part.} & \textbf{Random} \\
\midrule
Executed Gates & 105607 & 112480 & 131747 \\ 
\midrule
EPR Pairs & 361431 & 387522 & 459477 \\
\midrule
Magic State Travel & 35701 & 38047 & 40482  \\
\midrule
Cycles & 12324 & 13128 & 18378  \\ 
\bottomrule
\end{tabularx}
\label{table:7}
\end{table}

Table \ref{table:8} presents a benchmark involving a Cuccaro Adder circuit \cite{cuccaro2004newquantumripplecarryaddition}. In this circuit, CNOT, and T gates are dominant and almost symmetrically distributed among the qubits. Following core placement optimization, a significant improvement in decoding cycles and EPR pairs is observed relative to random mapping.

\begin{table}
\caption{Quantum Adder Circuit 12,117 Gate Circuit | 6x6 Mesh | 5x5 Core Size | Code Distance 3 | 468 Logical Qubits }
\centering
\small % make text slightly smaller to fit
\renewcommand{\arraystretch}{1.3}
\setlength{\tabcolsep}{4pt} % reduce horizontal padding
\begin{tabularx}{\columnwidth}{@{}p{4cm} >{\centering\arraybackslash}m{3cm} >{\centering\arraybackslash}X >{\centering\arraybackslash}X@{}}
\toprule
\textbf{Stats-Mapping} & \textbf{Part.+ QAPFA} & \textbf{Part.} & \textbf{Random} \\
\midrule
Executed Gates & 228096 & 230045 & 299010 \\ 
\midrule
EPR Pairs & 548856 & 553824 & 712053 \\
\midrule
Magic State Travel & 192548 & 193989 & 207971  \\
\midrule
Cycles & 38568 & 39546 & 87573  \\ 
\bottomrule
\end{tabularx}
\label{table:8}
\end{table}

Apart from these results, our framework also tracks statistics regarding the classical communication. Specifically, we count all the measurement results and circuit instructions that must be routed between cores. Tables \ref{table:9},\ref{table:10}, and \ref{table:11} report the total amount of classical bits to complete execution of the circuit. These bits are routed from core-to-core as illustrated in Figure \ref{fig:CNOT}.

\begin{table}
\caption{Random 10,000 Gate Circuit | 12x8 Mesh | 3x3 Core Size | Code Distance 3 | 480 Logical Qubits | 60/40 one-qubit gate/two-qubit gate.}
\centering
\small % make text slightly smaller to fit
\renewcommand{\arraystretch}{1.3}
\setlength{\tabcolsep}{4pt} % reduce horizontal padding
\begin{tabularx}{\columnwidth}{@{}p{4cm} >{\centering\arraybackslash}m{3cm} >{\centering\arraybackslash}X >{\centering\arraybackslash}X@{}}
\toprule
\textbf{Stats-Mapping} & \textbf{Part.+ QAPFA} & \textbf{Part.} & \textbf{Random} \\
\midrule
Traffic(bits) & 4906906 & 5250268 & 6205824 \\ 
\bottomrule
\end{tabularx}
\label{table:9}
\end{table}

\begin{table}
\caption{Random 10,000 Gate Circuit | 8x12 Mesh | 3x3 Core Size | Code Distance 3 | 480 Logical Qubits | 60/40 one-qubit gate/two-qubit gate.}
\centering
\small % make text slightly smaller to fit
\renewcommand{\arraystretch}{1.3}
\setlength{\tabcolsep}{4pt} % reduce horizontal padding
\begin{tabularx}{\columnwidth}{@{}p{4cm} >{\centering\arraybackslash}m{3cm} >{\centering\arraybackslash}X >{\centering\arraybackslash}X@{}}
\toprule
\textbf{Stats-Mapping} & \textbf{Part.+ QAPFA} & \textbf{Part.} & \textbf{Random} \\
\midrule
Traffic(bits) & 5499568 & 5776172 & 6766422 \\ 
\bottomrule
\end{tabularx}
\label{table:10}
\end{table}

These three benchmarks provide a useful comparison of layouts. Although the ratio of the total volume between the table layouts (number of gates $\times$ number of qubits) is similar, post-optimized traffic between Table \ref{table:9} and Table \ref{table:11} shows a great difference. The layout in Table \ref{table:11} results in better traffic-to-volume ratio, since the cores are larger and the need for inter-core operations is lower compared to the layout in Tables \ref{table:9} and \ref{table:10}.

\begin{table}
\caption{Random 10,000 Gate Circuit | 6x6 Mesh | 5x5 Core Size | Code Distance 3 | 468 Logical Qubits | 60/40 one-qubit gate/two-qubit.}
\centering
\small % make text slightly smaller to fit
\renewcommand{\arraystretch}{1.3}
\setlength{\tabcolsep}{4pt} % reduce horizontal padding
\begin{tabularx}{\columnwidth}{@{}p{4cm} >{\centering\arraybackslash}m{3cm} >{\centering\arraybackslash}X >{\centering\arraybackslash}X@{}}
\toprule
\textbf{Stats-Mapping} & \textbf{Part.+ QAPFA} & \textbf{Part.} & \textbf{Random} \\
\midrule
Traffic(bits) & 3982107 & 4124900 & 4716853 \\ 
\bottomrule
\end{tabularx}
\label{table:11}
\end{table}

\section{Conclusion and Future Work}
In this work, we constructed a system that can be used to understand the necessary resources for DQC with logical qubits. We have completed a three-phase process. First, we designed a scheduling algorithm that uses a network of ancilla qubits to route the logical states from one position to another. Then, we used a classical NoC to emulate the classical traffic for all these operations. This gives a clear understanding of classical resource consumptions and timing of a quantum circuit that use this system. Finally, we designed a robust mapping framework that takes care of minimizing the two-qubit gates between cores while also optimizing the total travel of magic states for the T and S gates. We think that this approach is wise for two points. Firstly, it reduces the routing of magic states for better quality T and S gates. Secondly, it places the qubits that are used for the T and S gates preferentially close to a point. This could bring advantages since these logical qubits should be decoded with a higher priority over other qubits. Essentially, the core idea is to approach the two-qubit gates and the T gates jointly. Such a system should reduce the inter-core CNOTs. At the same time, it should also give a higher priority to the qubits that have more T and S gates, since for decoding and better quality magic states, these qubits should be placed near the MSF and decoder units. 

This work could be extended by incorporating decoders and an additional optimization layer for their placement. Decoders are a crucial component of fault-tolerant quantum applications, and it would be valuable to investigate how they influence classical resource consumption as well as the compilation of quantum circuits. An interesting direction for another potential future work is to study the role of EPR pair distribution in this framework. Optimizing how entanglement is generated, allocated, and consumed could reveal important trade-offs between classical resource usage and quantum circuit compilation. Adding an optimization layer dedicated to entanglement management would provide significant insights and benefits to our proposed framework.

\section{Acknowledgments}
Authors acknowledge funding from the EC
through HORIZON-EIC-2022-PATHFINDEROPEN-01-101099697 (QUADRATURE).
\bibliographystyle{ACM-Reference-Format}
\bibliography{ref/ref}

\end{document}